\documentclass[prb,aps,twocolumn]{revtex4}
\usepackage{graphicx}
\usepackage{epstopdf}
\usepackage{dcolumn}
\usepackage{bm}
\usepackage{color}
\usepackage{ulem}
\usepackage{amsfonts,amsthm}
\usepackage{amsmath}
\usepackage{amssymb}

\begin{document}
\newcommand{\bR}{\mbox{\boldmath $R$}}
\newcommand{\tr}[1]{\textcolor{red}{#1}}
\newcommand{\trs}[1]{\textcolor{red}{\sout{#1}}}
\newcommand{\tb}[1]{\textcolor{red}{#1}}
\newcommand{\tbs}[1]{\textcolor{blue}{\sout{#1}}}
\newcommand{\tcyan}[1]{\textcolor{cyan}{#1}}
\newcommand{\Ha}{\mathcal{H}}
\newcommand{\mh}{\mathsf{h}}
\newcommand{\mA}{\mathsf{A}}
\newcommand{\mB}{\mathsf{B}}
\newcommand{\mC}{\mathsf{C}}
\newcommand{\mS}{\mathsf{S}}
\newcommand{\mU}{\mathsf{U}}
\newcommand{\mX}{\mathsf{X}}
\newcommand{\sP}{\mathcal{P}}
\newcommand{\sL}{\mathcal{L}}
\newcommand{\sO}{\mathcal{O}}
\newcommand{\la}{\langle}
\newcommand{\ra}{\rangle}
\newcommand{\ga}{\alpha}
\newcommand{\gb}{\beta}
\newcommand{\gc}{\gamma}
\newcommand{\gs}{\sigma}
\newcommand{\vk}{{\bm{k}}}
\newcommand{\vq}{{\bm{q}}}
\newcommand{\vR}{{\bm{R}}}
\newcommand{\vQ}{{\bm{Q}}}
\newcommand{\vga}{{\bm{\alpha}}}
\newcommand{\vgc}{{\bm{\gamma}}}
\newcommand{\mb}[1]{\mathbf{#1}}
\def\vec#1{\boldsymbol #1}
\arraycolsep=0.0em
\newcommand{\Ns}{N_{\text{s}}}
%

\title{
Asymmetric melting of one-third plateau in kagome quantum antiferromagnets
}

\author{
Takahiro Misawa$^1$, Yuichi Motoyama$^1$, and Youhei Yamaji$^{2,3}$
}

\affiliation{$^1$Institute for Solid State Physics, University of Tokyo, 5-1-5 Kashiwanoha, Kashiwa, Chiba 277-8581, Japan}
\affiliation{$^2${Department of Applied Physics}, University of Tokyo, Hongo, Bunkyo-ku, Tokyo, 113-8656, Japan}
\affiliation{$^3$JST, PRESTO, Hongo, Bunkyo-ku, Tokyo, 113-8656, Japan}

\date{\today}

\begin{abstract}
Asymmetric destruction of the one-third magnetization plateau upon heating is found 
in the spin-1/2 kagome Heisenberg antiferromagnets
using the typical pure quantum state approach.
The asymmetry originates from a larger density of states of low-lying excited states of $N_{\rm s}$ spin systems
with magnetization $(1/3-2/N_{\rm s})$ than that of low-lying states with magnetization $(1/3+2/N_{\rm s})$.
The enhanced specific heat and entropy that reflect the larger density of states in the lower-field side of the plateau
are detectable in candidate materials of the kagome antiferromagnets.
We discuss that the asymmetry
originates from the unprecedented preservation of the ice rule around the plateau.
\end{abstract}

\pacs{75.10.Jm, 75.30.Kz, 75.40.Cx}

\maketitle

\section{Introduction}
Interplay between the geometrical frustrations and
the quantum fluctuations
often prohibits spontaneous symmetry breakings and
induces quantum spin liquid states in many-body quantum spins.
The spin-1/2 antiferromagnet on the kagome lattice
is one of the promising candidates of the quantum spin liquids and
intensive theoretical studies have been done
~\cite{Jiang_PRL2008,Lu_PRB2011,White2011,Schollwock2012,Balents2012,Hotta2013,Lauchli_2016,Ran_PRL2007,Nakano_JPSJ2011,Becca2015,He_PRX2017}. 
Although recent highly-accurate {simulations} 
show that
the long-range magnetic order is absent in the kagome lattice,
it is still under hot debate what kind of
the quantum spin liquid (e.g. 
gapped $Z_2$ spin liquid~\cite{Jiang_PRL2008,Lu_PRB2011,White2011,Schollwock2012,Balents2012,Hotta2013,Lauchli_2016} 
or gapless $U(1)$ Dirac spin liquid~\cite{Ran_PRL2007,Nakano_JPSJ2011,Becca2015,He_PRX2017}) 
{is realized.}

The magnetization process of the kagome 
Heisenberg model has attracted much interest
{{because} the magnetization 
{plateaus}
under magnetic fields show exotic 
orders~\cite{Zhitomirsky_PRL2001,Hida2001,Zhitomirsky_PRB2004,Nakano2010,Hotta2013,Orus2016}.}
Recent theoretical calculations have shown that 
the one-third magnetization plateau appears 
{in the spin-1/2 kagome Heisenberg model}~\cite{Hida2001,Hotta2013,Capponi_PRB2013,Orus2016}. 
The detailed exact diagonalization study, however,
claims that the one-third plateau
is not conventional plateau by examining the
critical exponents of the magnetization process~\cite{Nakano2010,Sakai_PRB2011,Nakano2014}.
It is also an unresolved issue 
what kind of magnetic order
realizes in the one-third plateau because
different magnetic orders are suggested~\cite{Hotta2013,Orus2016}.
 
{{Because} the one-third magnetization plateau 
is characteristic of}
the candidate materials for the kagome antiferromagnets~\cite{Ishikawa2015},
it is an important issue to clarify how 
the finite-temperature effects stabilize or destabilize the one-third plateau.
{Although finite-temperature properties at zero magnetic field
have been studied 
by several numerical methods~\cite{Elstner_PRB1994,Misguich_2007EPJ,Sindzingre_2007,
Misguich_PRB2005,Nakamura_PRB1995,Sugiura_PRL2013,Shimokawa_JPSJ2016},
finite-temperature properties 
under magnetic fields such as the finite-temperature magnetization
are not systematically examined.}

In this {paper}, we study the thermodynamics of the
antiferromagnetic Heisenberg model on the kagome lattice 
under magnetic fields by using the 
TPQ state method~\cite{Sugiura2012} that enables us
to calculate thermodynamics of the quantum many-body systems
{including the kagome antiferromagnets~\cite{Sugiura_PRL2013,Shimokawa_JPSJ2016}}
in an unbiased way.
We note that similar methods were proposed 
in the pioneering works~\cite{Imada1986,Hams2000,FiniteLanczos}.
We {focus on}
the finite-temperature effects around the 
one-third magnetization plateau.

As a result, we have found that the peculiar asymmetric
collapse of the one-third plateau occurs at finite temperatures.
This asymmetric collapse can be explained by 
the large degeneracy existing just below the plateau.
We will discuss that the asymmetric degeneracy can be
explained by the preservation of the ice rule.
In the one-third plateau,
we have {shown} 
that the $\sqrt{3}\times\sqrt{3}$
magnetic order {is realized} at the ground state but
several apparent disordered states compete 
with the $\sqrt{3}\times\sqrt{3}$ order,
which is consistent with previous studies~\cite{Honecker_PhysicaB2005,Cabra_PRB2005,Orus2016}.
Reflecting the anomalous degeneracies,
the entropy remains large even at the plateau down to 
the lowest temperatures.
We have also found that the
enhancement of the entropy and the specific heat 
just below the plateau appears,
which can be detected experimentally.

This paper is organized as follows:
In Sec. II, we introduce the antiferromagnetic Heisenberg model
on the kagome lattice and explains the details of the TPQ method.
In Sec. III, we show the results of the finite-temperature 
effects on the one-third magnetization plateau.
In Sec. IV, we analyze the low-energy excited states around the
one-third magnetization plateau by 
using the locally optimal block conjugate gradient
(LOBCG) method. 
In Sec. V, we show results of the magnetic field dependence of the
thermodynamic properties such as the entropy and the specific heat.
Sec. VI is devoted to the summary and discussions.
\section{Model and Methods}
We study the antiferromagnetic Heisenberg model on the kagome lattice 
defined as
\begin{align}
{\cal H} &= 
J\sum_{\langle i,j\rangle}\vec{S}_{i}\cdot\vec{S}_{j}-h\sum_{i}S_{i}^{z}
\label{eq:H}
\end{align}
where 
$\vec{S}_{i}$ is a spin operator of the localized {spin-1/2} at $i$th site
and $h$ denotes the magnetic field. 
The antiferromagnetic interactions $J>0$
only exist between the nearest-neighbor sites $\langle i,j\rangle$
on the kagome lattice.
The magnetization $m$ is defined as $m=\frac{1}{SN_{\rm s}}\sum_{i}\langle S_{i}^{z}\rangle$,
where $N_{\rm s}$ is the number of the sites and 
we take $S=1/2$.
In this definition, saturation magnetization is given by $m_{\rm s}=1$.
We mainly analyze {27 site} and {36 site} clusters for the kagome lattice
with a periodic boundary condition in the present {paper}.
For comparison, we perform the calculations for 
the triangular lattice ({27 site} clusters), 
which also has the one-third plateau under magnetic fields.

\begin{figure}[b!]
  \begin{center}
    \includegraphics[width=8cm,clip]{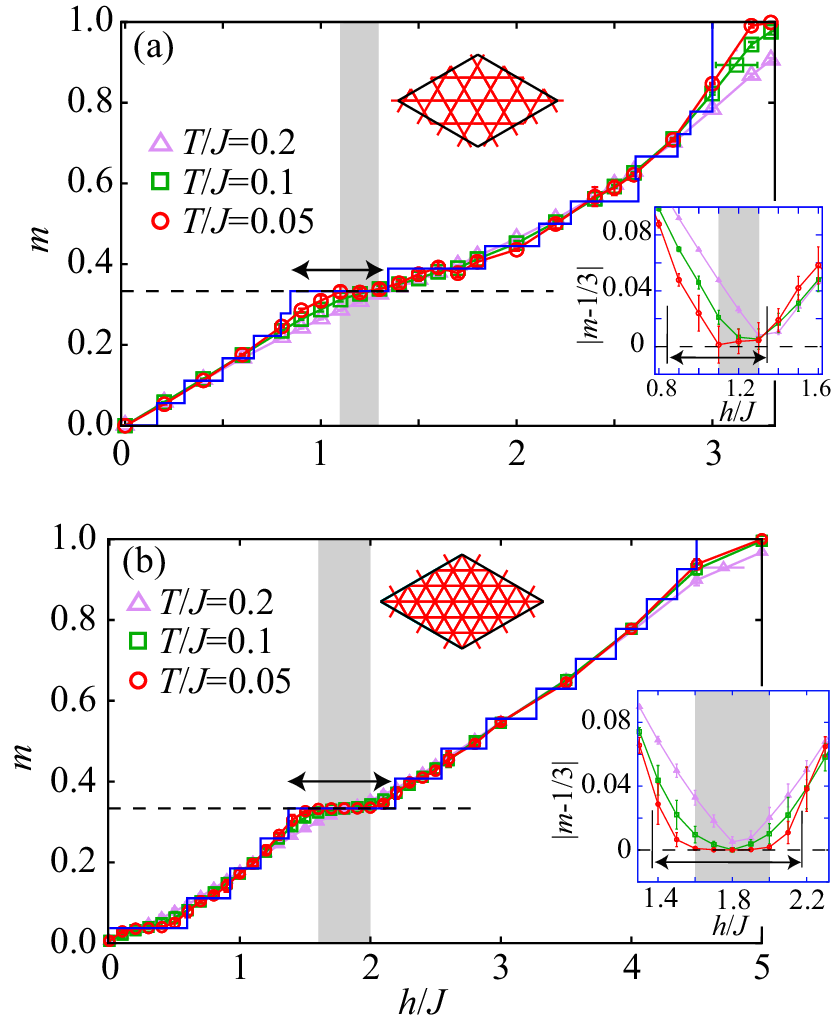}
  \end{center}
\caption{(color online).~Finite-temperature magnetization process of the 
{spin-1/2} antiferromagnetic Heisenberg model
(a) on the kagome lattice and (b) the triangular lattice.
{The magnetization at $T=0$ is also shown by a solid line.}
In the {left insets}, the lattice structures 
for {36 site} kagome lattice and {27 site} triangular lattice are shown.
The one-third plateau region {($m=1/3$ within error bars)} at $T/J=0.05$ 
is shown by the shaded region.
The arrow shows the region where the one-third
plateau appears at zero temperature for $N_{\rm s}=36$ (kagome lattice)
and $N_{\rm s}=27$  (triangular lattice).
{In the {right insets}, we show $|m-1/3|$ around the 
plateaus for both lattices.}}
\label{fig:mag}
\end{figure}

In this {paper}, we use the TPQ state
to analyze the finite-temperature properties.
In the TPQ method, we {iteratively} generate
the $k$th TPQ state as
\begin{align}
|\psi_{k}\rangle\equiv\frac{(l-\hat{\cal{H}}/N_{\rm s})|\psi_{k-1}\rangle}{|(l-\hat{\cal{H}}/N_{\rm s})|\psi_{k-1}\rangle|},
\end{align}
where $|\psi_{0}\rangle$ is an initial random wavefunction 
and $l$ is a constant that is larger than
the maximum eigenvalue of ${\cal{H}}/N_{\rm s}${~\cite{Sugiura2012}}.
In this study, we typically take $l/J=3$.
From the $k$th TPQ state,
we can calculate thermodynamic properties such as 
temperature, internal energy, 
and the spin correlations~\cite{Sugiura2012,hphi}.
To examine properties of the ground state and low-energy excited states, 
we also perform 
the LOBCG method~\cite{LOBCG}.

{Here, we mention about the limitation of the TPQ method.
Although the TPQ method gives the unbiased and essentially exact results
for given finite system sizes, the errors of the TPQ method, 
{which are defined by the standard deviations of the statistical distribution of the initial
random vectors $|\psi_{0}\rangle$},
is bounded by the residual entropy~\cite{Sugiura2012,Sugiura_PRL2013}. 
If the residual entropy is large, 
the {errors} of the TPQ method become smaller, i.e., 
the initial vector dependence of the
physical quantities becomes small. 
Thus, 
in the previous works~\cite{Sugiura_PRL2013,Yamaji_PRL2014,Shimokawa_JPSJ2016,Misawa_JPSJ2018,Laurell_npj2020}, 
the TPQ method is used for examining the finite-temperature properties
of the quantum spin liquid where 
the large remaining entropy is expected.
As shown in Appendix A, for s small system size,
we confirm that the TPQ method 
well reproduces the result obtained by the full diagonalization 
(see Fig.~\ref{fig:tpq_ene}).}
{We note that the average values of the 
physical quantities become exact in the limit
of a large number of samplings 
when we independently evaluate the 
averages of the distribution function and the physical 
quantities, as is done in the canonical TPQ method~\cite{Sugiura_PRL2013}.}

{
In the following calculations,
we evaluate the error bars of the TPQ calculations by choosing
several independent initial vectors. As we show later, in the relevant temperature
region ($T\geq0.05$), the errors of the TPQ calculations are small.
We can safely discuss the finite-temperature effects on the 1/3 plateau
in the kagome lattice.}

\section{Finite-temperature effects on the one-third magnetization plateau}
In Fig.~1(a) and (b), we show the finite-temperature magnetization process 
of the Heisenberg model 
on the kagome lattice and the triangular lattice.
To see the finite-temperature effects on the one-third plateaus,
we show the width of the one-third 
plateaus at zero temperature as arrows
for both lattices.

At high temperature ($T/J=0.2$),
the magnetization changes smoothly
as a function of the magnetic field for both lattices.
By lowering the temperature, 
a plateau-like structure appears around
$h/J=1.2$ in the kagome lattice 
and $h/J=1.8$ in the triangular lattice {(see the right insets in Fig.1)}.
We, however, find that 
the width of the plateau in the kagome lattice is 
largely different from that of the zero-temperature limit 
even at the lowest temperature ($T/J=0.05$), 
while the width of the plateau in the triangular lattice 
seems to smoothly converge to the zero-temperature limit.
The asymmetric behavior in the
kagome lattice indicates that 
the low-energy {excitations} 
above and below the one-third plateau {are} different, i.e.,
the density of states for {the lower-field} side is
larger than
the density of states for {the higher-field side}.
We will confirm that the asymmetric structure
exists in the low-energy excited states.
{We note that error bars around the 1/3 plateau in the 
kagome lattice are large, and this may originate from the
peculiar degeneracy in the 1/3 plateau, as we will discuss later.}

In Fig.~2(a), we show that the 
temperature dependence of the magnetization in the kagome lattice
for several different magnetic fields.
Dashed vertical lines show the temperatures
where we plot the magnetization process in Fig.~1(a).
{When the temperature is decreased},
the magnetization 
converges to $m=1/3$
{around $h/J=1.2$, which}
indicates the formation
of the plateau at the finite temperature. {As shown in Fig.~2(b),
we confirm that the magnetization converges
to $m=1/3$ at low temperatures around $h/J=1.2$.}

As we mentioned above, around $h/J=0.9$, 
the plateau appears at 
zero temperature while it vanishes at finite temperatures ($T/J=0.05$).
As shown in Fig.~2(c), because we find that finite-size effects are small
at $h/J=0.9$, it is plausible that the 
{fragility} of the magnetization
plateau at a low magnetic field side is intrinsic phenomena.

{In addition to the one-third plateau around $h/J=1.2$,
we find that the apparent magnetization plateau ($m=14/36$) exists
slightly above the one-third plateau, i.e.,
around $h/J=1.6$ as shown in Figs.~2(a) and (c).
{Because} the system size dependence of the 
apparent plateau is large, as shown in Fig.~2(c), 
the additional plateau seems to be an
artifact of the finite-size effects
and may vanish in the thermodynamic limit.
The apparent magnetization originates from the
``ramp'' structure observed in zero-temperature calculations,
where the magnetization smoothly converges to the plateau above the
one-third plateau~\cite{Nakano2010}.
In finite-size systems, this ramp structure induces 
the apparent plateau due to the discreteness of the magnetization.}

\begin{figure}[htb!]
  \begin{center}
    \includegraphics[width=8cm,clip]{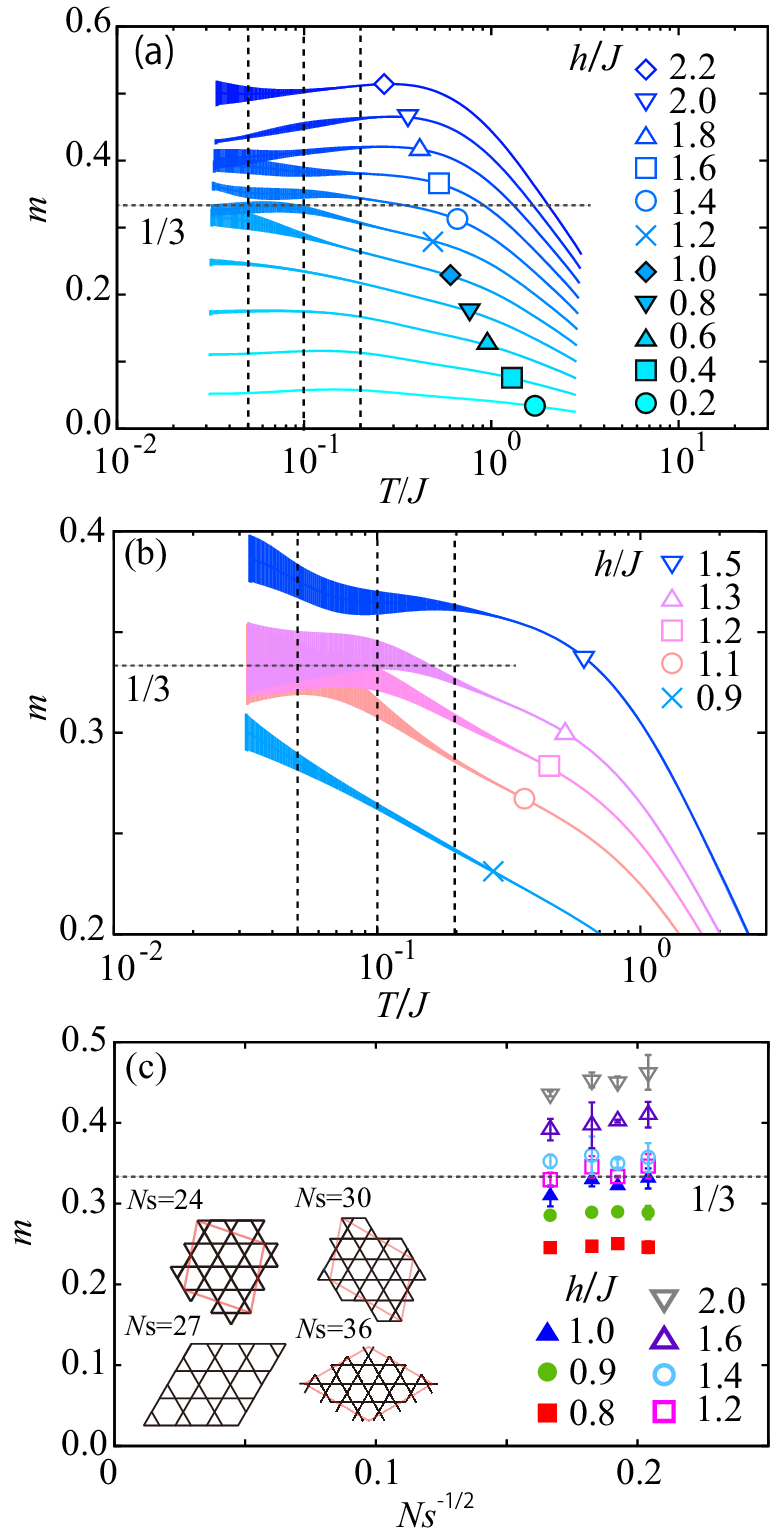}
  \end{center}
\caption{(color online).
~(a)~Temperature dependence
of the magnetization in the kagome lattice ({36 site} clusters) 
for several different magnetic fields.
Around $h/J=1.2$, magnetization saturates to $1/3$.
Slightly above the one-third plateau,
we also find the signature of 
another apparent plateau around $h/J=1.6$.
However, we conclude that this plateau is induced by the
finite-size effects as we discussed in the main text.
(b)~{Temperature dependence of the magnetization around
the $1/3$ plateau. For $h/J=1.1,1.2,$ and $1.3$,
the magnetizations converge to 1/3 at low temperatures.}
(c)~Size-dependence of the magnetization at $T/J=0.05$ 
for different magnetic fields. 
We take $N_{\rm s}=24,27,30$, and $36$ clusters.
In the {insets}, the shapes of the used clusters are shown.
{The size-dependent error bars around 1/3 plateau
may originate from the peculiar degeneracy in the 1/3 plateau.}
}
\label{fig:temp}
\end{figure}

\section{Low-energy excitations around the one-third plateau}
Here, we examine the low-energy {excitations} around the plateau
by using LOBCG method~\cite{LOBCG}. 
In the inset of Fig.~3(a), 
we show the lowest 128 eigenvalues for 
the {27 site} cluster and the lowest 16 eigenenergies for
the {36 site} cluster around the plateau.
Irrespective of the system sizes, we find 
a large density of states exists
just below the plateau, i.e., $m=7/27$ ($m=10/36$) for the 27 (36) {site} clusters.
{We define the density of states  as the {number of eigenstates per unit energy $J$}.
{As we show, we confirm that the slope of the magnetization curve (susceptibility) 
is proportional to the density of states, i.e., $\frac{\partial m}{\partial h}\propto {\rm(density~of~states)}$}}
For the {27 site} clusters, all the 128 eigenenergies
exist within $0.2J$ from the ground states.
In sharp contrast to this,
{the density of states} 
is small just above the plateau
[$m=11/27$ (27 sites) and $m=14/36$ (36 sites)].
{We note that the large density of states
is consistent with the anomalous 
steep magnetization process just below the plateau,
which is pointed out 
in the previous studies~\cite{Nakano2010,Sakai_PRB2011,Nakano2014}.}
{This can be rephrased as the difference of the
degeneracy induces the ramp structure
in the magnetization process~\cite{Nakano2010}, i.e., 
the large degeneracy 
below the plateau induces the steep change in the magnetization
and small degeneracy above the plateau induces
the small {slope in magnetization process}.}

\begin{figure}[t!]
  \begin{center}
    \includegraphics[width=8cm,clip]{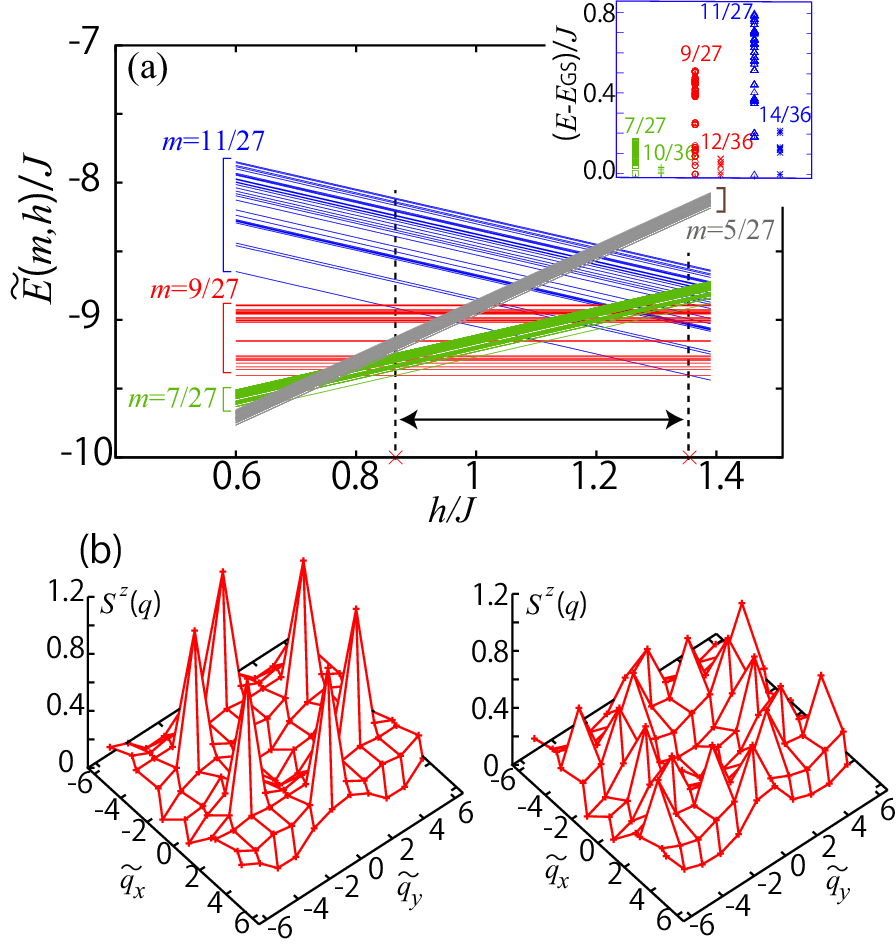}
  \end{center}
\caption{(color online).~
(a)~Low-energy spectrum around the one-third magnetization plateau 
for the {27 site} cluster. 
We plot $\tilde{E}(m,h)=E(m)-h\times (m-1/3)$ for $m=7/27,9/27=1/3,11/27$.
Dotted lines show the position of the crossing points
of the lowest energies.
The arrow indicates the width of the plateau at zero temperature.
In the inset,
we show the lowest 128 eigenenergies for {27 site}
clusters ($m={5/27},7/27,9/27,11/27$) and 
the lowest 16 eigenenergies 
for {36 site} clusters ($m=10/36,12/36,14/36$).
The eigenenergies are measured from the ground-state energy $E_{\rm GS}$.
(b)~Spin structure factors 
in the extended Brillouin zone
for the {36 site} cluster at $m=1/3$ for
the ground state (left panel) and 
the third excited state (right panel), where
$\tilde{q}_{\alpha}=q_{\alpha}\times\frac{3}{\pi}~(\alpha=x,y)$.
In the ground state, there are sharp peaks at
$\vec{q}=(\pm4\pi/3,0),(\pm2\pi/3,\pm2\pi/\sqrt{3})$,
which correspond to the $\sqrt{3}\times\sqrt{3}$ order.
In contrast to that,
we find no significant peaks in the spin structure
factor in the third excited state whose
energy difference is given by $E_{\rm third}-E_{\rm GS}\sim0.04J$.
}
\label{fig:spec}
\end{figure}

{To see how the asymmetric low-energy {excitations}
affect 
the
finite-temperature magnetization process,
we plot the magnetic field dependence of the 
low-energy spectrum $\tilde{E}(m,h)=E(m)-h(m-1/3)$
in Fig.~3(a).
Because the degeneracy between different 
$m$ sectors, i.e., $m=1/3$ and $m=7/27$,
exists around the lower endpoint of 
the plateau $h/J\sim0.8$,
the different magnetizations 
can be easily mixed by the finite-temperature effects.
{We also point out that $m=5/27$ can largely affect
the lower bound of the plateau as shown in Fig.~3(a).}
This is the reason why the plateau is destroyed 
around $h/J\leq0.9$ even at the lowest temperature $T/J=0.05$.
In contrast to that,
around the higher endpoint of the plateau ($h/J\sim1.2$), 
the degeneracy is weak and 
the magnetization plateau is robust
against the finite-temperature effects.}

In Fig.~3(b),
we show the spin structure factors $S^{z}(\vec{q})$, 
which is defined as
\begin{align}
S^{z}(\vec{q})=\frac{1}{N_{\rm s}}\sum_{i,j} e^{{\rm i}\vec{q}(\vec{r}_{i}-\vec{r}_{j})}(S_{i}^{z}-\langle S_{i}^{z}\rangle)(S_{j}^{z}-\langle S_{j}^{z}\rangle),
\end{align}
where $r_{i}$ denotes the position of the
lattice point in the two dimensions.
We find that spin structure factors have sharp 
peaks at $\vec{q}=(\pm4\pi/3,0),(\pm2\pi/3,\pm2\pi/\sqrt{3})$
{in the ground state.}
{This indicates that the 
$\sqrt{3}\times\sqrt{3}$ spin structures realize in 
the ground state.
We also find that the several excited states
that do not show any signature of magnetically ordered states
compete with the $\sqrt{3}\times\sqrt{3}$ orders
(see the right panel in Fig.~3(b)).}
{One of the candidates of the 
apparent disordered state is the nematic state.
Further detailed analysis of the nature of
the disordered state is left for future studies.}
This result indicates that degeneracy still exists 
even in the plateau region~\cite{Cabra_PRB2005,Orus2016}.
As we show later,
this degeneracy induces the {large} remaining entropy
at low temperature.

\section{Magnetic-field dependence of entropy and specific heat}
In Fig.~4(a),
we show the temperature dependence 
of the {specific heat} $C$, which is
defined as 
\begin{align}
C=\frac{\langle H^{2}\rangle-\langle H\rangle^2}{N_{\rm s}T^2}.
\end{align}
{At $h=0$, 
we obtain the same temperature dependence of the 
{specific heat} in the previous studies~\cite{Sugiura_PRL2013,Shimokawa_JPSJ2016}.
We find that the shoulder structure at $h/J=0$ below $T/J\sim0.2$, 
which roughly corresponds {to} the singlet-triplet 
excitation gap~\cite{Waldtmann_EP1998,Nakano_JPSJ2011},
is immediately vanished by applying the magnetic field ($h/J=0.2$).}
Thus, as shown in Fig.~4(b), 
the entropy at finite temperatures ({$T/J\leq0.1$}),
which is  defined as
\begin{align}
  {S(T)=1-\frac{1}{\rm ln2}\int_{T}^{\infty}\frac{C}{T}\mathrm{d}T},
\end{align}
increases
{when}
the magnetic 
fields are applied. 
In the plateau region,
the entropy is relatively low but 
the entropy is not fully released  
down to the lowest temperature $T/J=0.05$.
This result {is} consistent with the fact that
the degeneracy still exists in the plateau.

{We show the magnetic-field dependence of the
specific heat {coefficient} $C/T$ in Fig.~4(c).
Just below the one-third plateau,
we find the enhancement of $C/T$, which is 
consistent with the degeneracy 
existing in the low-energy excited state.
This large enhancement of $C/T$ is the characteristic feature
of the plateau in the kagome lattice and
is expected to be detected in experiments.
We note that the behaviors of the {specific heat} 
around the saturation field ({peaks and dip around $h/J=3$}) are consistent 
with those of the hard hexagon model~\cite{Zhitomirsky_PRB2004}, 
which well describes the
low-energy degrees of freedom of 
the quantum kagome Heisenberg model
around the saturation field.}

{Here, we clarify the origin 
of the {\it asymmetric plateau melting} and 
the difference in the density of states for $m =1/3-2/N_{\rm s}$ and $m = 1/3 + 2/N_{\rm s}$
in the kagome-lattice Heisenberg model. 
The asymmetry originates from the ice-rule configuration of
spins around the one-third plateau.
As shown in the inset of Fig.~4(b), 
the ground-state and low-lying excited-state 
wave functions show a large probability 
of adhering to the ice rule at $m=1/3$ for the kagome lattice while
the probability is relatively low for the triangular lattice. 
Therefore, we can assume
that the spin configurations 
for $m\sim 1/3$ are generated by keeping the 
ice-rule constraint as much as possible.
Under the constraint, there are $N_{\rm s}/3$ ($2N_{\rm s}/3$) flippable spins 
to increment (decrement) the total $S^z$ from the one-third plateau by 1.
This asymmetry is the origin of the asymmetric plateau melting. }
{The degeneracy lifting by the kinetic energy of a defect (flipped spin) in the ice manifold
for $m=1/3 + 2/N_{\rm s}$ may explain the smaller entropy for $m>1/3$ 
than that for the $m=1/3$ state.
{We note that the preservation of the ice-rule is 
consistent with the fact that 
the one-third plateau at the Heisenberg point 
has a large overlap with its Ising limit~\cite{Cabra_PRB2005}.}
}

\begin{figure}[htb!]
  \begin{center}
    \includegraphics[width=8cm,clip]{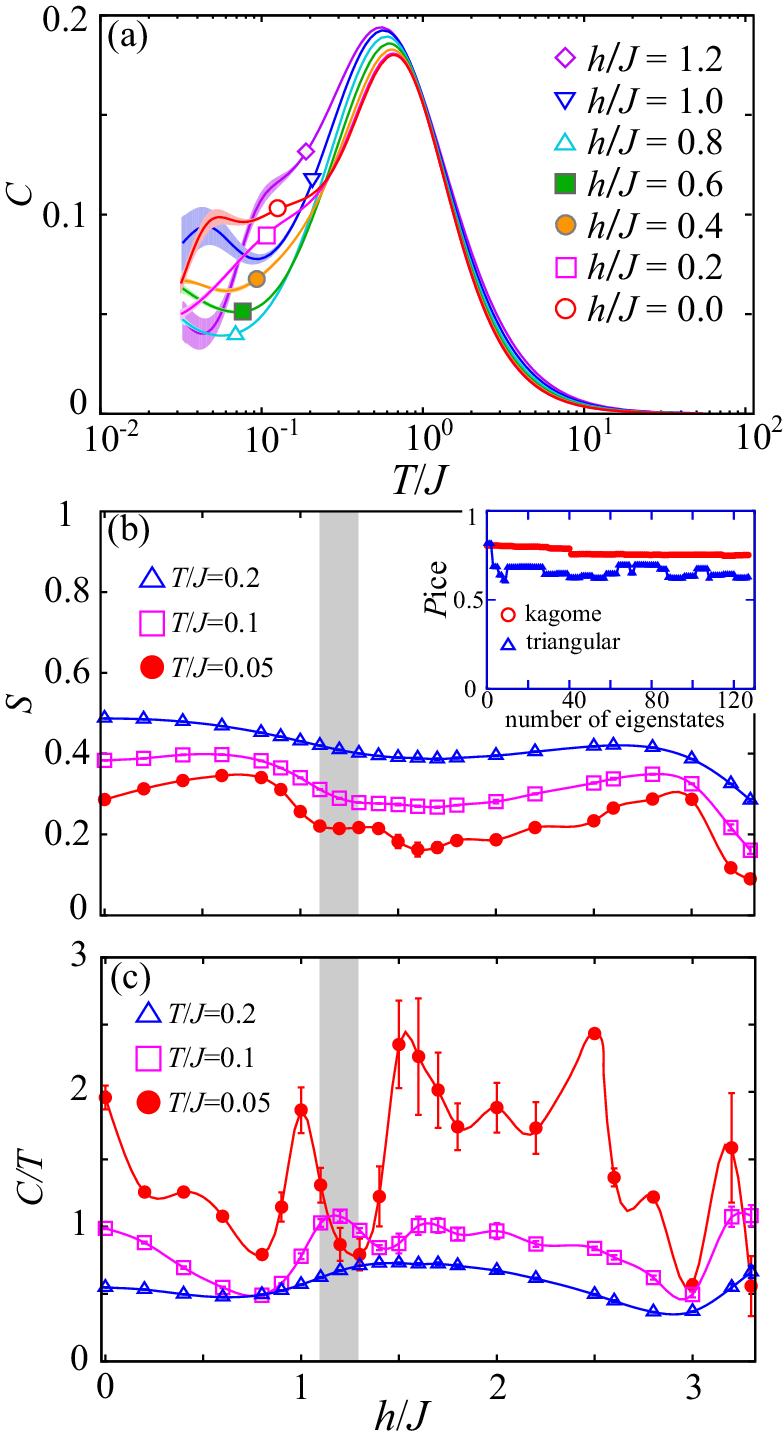}
  \end{center}
\caption{(color online).~
(a)~Temperature dependence
of the specific heat for several different magnetic fields.
(b)~Magnetic field dependence of the entropy $S$ for
several different temperatures.
The region where the one-third plateau appears at $T/J=0.05$
is represented by the shaded region.
{In the inset, we show the probability of the ice rule $P_{\rm ice}$
for 128 eigenstates in {27 site} clusters.
$P_{\rm ice}$ is defined in Appendix B.
}
(c)~Magnetic-field dependence of the specific heat
$C/T$ for several different temperatures. 
At $T/t=0.05$, we find that 
$C/T$ has peaks at both the 
sides of the one-third plateau.
We note that the finite-size effects are small around the 
one-third plateau for both $S$ and $C/T$.
Solid curves are guides for eyes.
}
\label{fig:gamma}
\end{figure}

\section{Summary and Discussions}
In summary,
we analyze the thermodynamic properties of the spin-1/2 antiferromagnetic
Heisenberg model on the kagome lattice by using the TPQ method.
We show that the one-third magnetization plateau in the kagome lattice
is asymmetrically destroyed at finite temperatures.

By using the LOBCG method,
we show that the large degeneracy exists
just below the plateau. 
The degeneracy induces the asymmetric behavior and
the enhancement of both the entropy and the specific heat.
We also identify that the origin of the degeneracy 
is the unexpected robustness of the ice-rule configurations
at the plateau.
Our detailed and unbiased analyses on the thermodynamics of the
ideal quantum kagome antiferromagnets under magnetic fields
offer a firm basis for 
characterizing the experimental candidates 
of the kagome antiferromagnets~\cite{Ishikawa2015,Hiroi_JPSJ2001,Shores_JACS2005,Mendels_JPSJ2010,Okamoto_JPSJ2009,Okamoto_PRB2011}.

{
Recently, it is shown that we can treat
up to 50-site systems by reducing the
size of Hilbert dimensions using 
the symmetries inherent in systems~\cite{Wietek_PRE2018}.
By taking such large system size, we can
obtain a more definitive conclusion on the 
anomalous thermodynamic properties on the 1/3 plateau.
{We note that it is also a challenging
problem to study the temperature dependence of the static and 
dynamical spin correlation function based on the TPQ method~\cite{Yamaji_2018FT}.}
Further investigations along this direction are intriguing 
but left for future studies.
}

{\it Note added.}---
{Recently, we became aware of 
independent numerical work~\cite{Chen_2018} 
that treats the finite-temperature properties of the
kagome-lattice Heisenberg model
but focuses on a different aspect.}
We also became aware that the similar asymmetric melting was independently reported in 
Ref.~[\onlinecite{Schnack_PRB2018}] after submitting the article.

\begin{acknowledgments}
A part of calculations is done by using open-source software $\mathcal{H}\Phi$~\cite{hphi,hphi_ma,hphi_git}.
Our calculation was partly carried out at the 
Supercomputer Center, Institute for Solid State Physics, University of Tokyo.
This work was supported by JSPS KAKENHI (Grant Nos.~16K17746 and 16H06345)
and was supported by PRESTO, JST (JPMJPR15NF).
This work was also supported in part by MEXT as a social 
and scientific priority issue (Creation of new functional devices and high-performance materials
to support next-generation industries) to be tackled by using post-K computer.
TM and YM acknowledge Tsuyoshi Okubo and Naoki Kawashima for
useful discussions.  
TM and YM were supported by
Building of Consortia for the Development of Human Resources
in Science and Technology from the MEXT of Japan.
\end{acknowledgments}

\appendix
\clearpage

\section{Comparison with full diagonalization}
\begin{figure}[htb!]
  \begin{center}
    \includegraphics[width=8cm,clip]{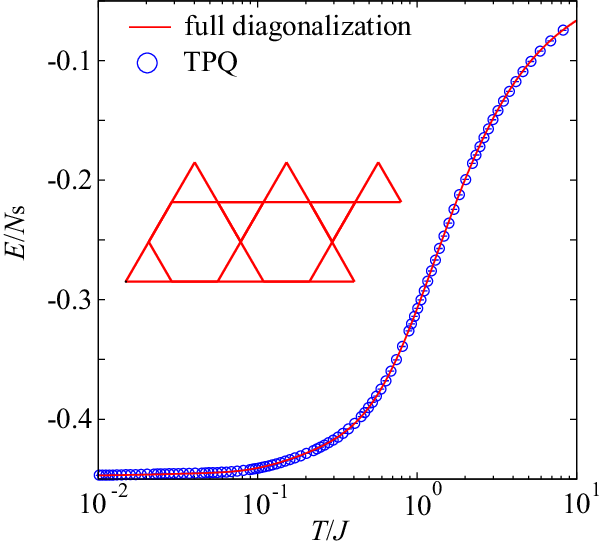}
  \end{center}
\caption{(color online).~Temperature dependence of the
energy obtained by the TPQ method (blue circles) and
the full diagonalization (red curve).
For the TPQ calculations,
we perform 5 independent runs and regard its standard deviation as error bars.
In the inset, we show the geometry used in the calculation.
}
\label{fig:tpq_ene}
\end{figure}

Here, we examine the accuracy of the TPQ method
by comparing with the result obtained 
by the full diagonalization.
In Fig.~\ref{fig:tpq_ene},
for 18-site kagome lattice with total $S_{z}=0$ 
(dimension of the Hilbert space is 48620),
we show the temperature dependence of the energy
obtain by the TPQ method and the full diagonalization.
From this result, we confirm that the
TPQ method well reproduces exact results.

\section{Definitions of ice rule}
In this appendix, 
we show the definitions of the ice rule.
We defined the local measure of the ice rule on one upper triangular defined as
\begin{align}
n_\triangle
&=
\frac{1}{3}\left\{
\left(S^z_{\triangle(1)}-\frac{1}{2}\right)
\left(S^z_{\triangle(2)}-\frac{1}{2}\right)
\left(S^z_{\triangle(3)}+\frac{1}{2}\right)
\right. \\
&\quad+
\left(S^z_{\triangle(1)}-\frac{1}{2}\right)
\left(S^z_{\triangle(2)}+\frac{1}{2}\right)
\left(S^z_{\triangle(3)}-\frac{1}{2}\right)
\\
&\quad+
\left.
\left(S^z_{\triangle(1)}+\frac{1}{2}\right)
\left(S^z_{\triangle(2)}-\frac{1}{2}\right)
\left(S^z_{\triangle(3)}-\frac{1}{2}\right)
\right\} \\ \notag
&=
S^z_{\triangle(1)}S^z_{\triangle(2)}S^z_{\triangle(3)}
\\
&\quad
-\frac{1}{6}\left(
S^z_{\triangle(1)}S^z_{\triangle(2)}
+
S^z_{\triangle(2)}S^z_{\triangle(3)}
+
S^z_{\triangle(3)}S^z_{\triangle(1)}
\right)
\\ \notag
&\quad
-\frac{1}{12}\left(
S^z_{\triangle(1)}
+ S^z_{\triangle(2)}
+ S^z_{\triangle(3)}
\right)
\\ \notag
&\quad
+ \frac{1}{8},
\end{align}
where $S^z_{\triangle(i)}$ is the $z$ component of the spin operator
$S^{z}$ at the $i$th site of the upper triangle.
We note how to index the three sites on the triangular does not affect because these sites are equivalent.

By using this $n_{\triangle}$,
we calculate $P_{\rm ice}$
as follows:
\begin{align}
\Delta     &= \sum_{\triangle}n_{\triangle}, \\
P_{\rm ice}&= \frac{1}{N_{\rm norm}}\sum_{i}|a_{i}|^2\langle i|\Delta |i\rangle,
\end{align}
where $a_{i}$ is a coefficient of the $i$th real-space configuration $|i\rangle$,
$\Delta$ is number of upward triangles that satisfy the ice rule, and
$N_{\rm norm}$ is defined for $P_{\rm ice}$ becomes one when the ice rule
is completely satisfied.
We note that $P_{\rm ice}\sim 0.4707$ ($N_{\rm s}=27$) for a random configuration of spins
irrespective of the lattice geometry.


\end{document}